\documentclass[prd,twocolumn,nofootinbib,showpacs]{revtex4-1}

\usepackage{amsfonts}
\usepackage{amsmath}
\usepackage{amssymb}
\usepackage{bm}
\usepackage{dcolumn}
\usepackage{graphicx}
\usepackage{graphics}
\usepackage[latin1]{inputenc}
\usepackage{latexsym}
\usepackage{rotating}
\usepackage[colorlinks=true]{hyperref}
\usepackage[all]{hypcap} 
\usepackage{xspace} 
\usepackage[usenames]{color}
\usepackage{mathrsfs}

\usepackage{ulem}
\definecolor {darkgreen}{rgb}{0.2,0.7,0.2}

\newcommand\be{\begin{equation}}
\newcommand\ba{\begin{eqnarray}}
\newcommand\ee{\end{equation}}
\newcommand\ea{\end{eqnarray}}

\newcommand\bw{\begin{widetext}}
\newcommand\ew{\end{widetext}}

\newcommand{\NS}{*}

\newcommand{\BH}{{\mbox{\tiny BH}}}
\newcommand{\BL}{{\mbox{\tiny BL}}}

\begin{document}
\title{Relating Follicly-Challenged Compact Stars to Bald Black Holes: \\
A Link between Two No-hair Properties}

\author{Kent Yagi}
\author{Nicol\'as Yunes}
\affiliation{Department of Physics, Montana State University, Bozeman, MT 59717, USA.}

\date{\today}

\begin{abstract} 

Compact stars satisfy certain no-hair relations through which their multipole moments are given by their mass, spin and quadrupole moment. 
These relations are approximately independent of their equation of state, relating pressure to density. 
Such relations are similar to the black hole no-hair theorems, but these possess event horizons inside which information that led to their formation can hide.
Compact stars do not possess horizons, so whether their no-hair relations are related to the black hole ones is unclear. 
We investigate how the two relations are related by studying relations among multipole moments for compact stars with anisotropic pressure as a toy model, which allows such stars to be more compact than those with isotropic pressure.
We here show numerically that the compact star no-hair relations approach the black hole ones as the compactness approaches that of a black hole. 
We also prove analytically that the current dipole moment exactly reaches the black hole limit quadratically in compactness as strongly-anisotropic stars approach the black hole limit.
We moreover show that compact stars become progressively oblate in this limit, even if prolate at low compactness due to strong anisotropies. 

\end{abstract}



\maketitle

\allowdisplaybreaks{}

\section{Introduction}

\emph{The Hair of Black Holes.}~Astrophysical black holes are said to have \emph{no hair} because their exterior gravitational field can be completely described by only two \emph{observable} quantities: their mass and spin angular momentum. All other information or \emph{hair} that may have led to the formation of the black hole is hidden inside its event horizon. Electric charge is a third piece of information, but this is typically neglected since astrophysically realistic black holes are expected to be approximately neutral.

The baldness of black holes and General Relativity's equivalence principle~\cite{will-living} imply that 
the motion of \emph{test} bodies around a black hole depends only on its mass and spin. This is of critical importance in astrophysics because it allows for the modeling of complicated systems irrespective of the details of the central object. For example, the motion of stars around the black hole at the center of the Milky Way~\cite{Liu:2011ae,Will:2007pp} can be modeled with only knowledge of the central object's mass and spin. 

One way to mathematically formalize the black hole no-hair relations is by inter-relating the \emph{multipole moments} of their exterior gravitational field. Multipole expansions are commonly employed in electrodynamics and in gravitational physics to represent fields outside and far from a localized source. For example, we can approximate the gravitational potential of a bounded source distribution as a series expansion in inverse powers of the distance, $r$, from the field point to the source. The coefficient $M_{\ell}$ that multiplies $(1/r)^{1+\ell}$ and spherical harmonics $Y_{\ell m}(\theta, \phi)$ is called the \emph{mass $\ell$th-pole moment}, e.g.~$M_{0}$ is the mass monopole, $M_{2}$ is the mass quadrupole, etc. In addition to mass moments, General Relativity requires  \emph{current multipole moments} $S_{\ell}$ to fully describe the gravitational field, as in electrodynamics when performing a multipolar decomposition of the vector potential. In gravity, this stems from the need to decompose the field generated by \emph{gravitational currents}, induced either by the motion of bodies or their spin~\cite{MTW}. 

The no-hair theorems for black holes then imply that, in the multipole expansion of their exterior gravitational field, only 2 multipole moments are independent and all others can be expressed in terms of these two. The independent ones for black holes are the mass monopole, $M_{0}(=M)$, and the current dipole, $S_{1} (=|\vec{S}| = S)$. All other multipoles (of order $\ell \geq 2$) are given by~\cite{carter-uniqueness,hansen}
\be
\label{eq:no-hair}
M_{\ell}^{\BH} + i S_{\ell}^{\BH} = M \left(i a \right)^{\ell}\,,
\ee
with $a \equiv S/M$. Thus, for example, we have that $(M_{0}^{\BH},S_{0}^{\BH}) = (M,0)$, $(M_{1}^{\BH},S_{1}^{\BH}) = (0,S)$, $(M_{2}^{\BH}$, $S_{2}^{\BH}) = (-S^{2}/M,0)$, $(M_{3}^{\BH},S_{3}^{\BH}) = (0,- S^{3}/M^{2})$. Notice that the ratio $a/M$ can be close to unity in Nature, and hence, higher multipole moments need not be small.

These astonishing results were proven analytically in the late '60s and '70s by Robinson~\cite{robinson}, Israel~\cite{israel,israel2}, Hawking~\cite{hawking-uniqueness0,hawking-uniqueness}, and Carter~\cite{carter-uniqueness}. Their work relied on a few assumptions, such as that black holes are vacuum solutions to the Einstein equations. Non-vacuum solutions, such as those that represent stars, evade the no-hair theorems. Therefore, the multipole expansion of the exterior gravitational field of stars should depend on an infinite number of \emph{independent} multipole moments. 

The above implies that multipole moments of higher order than $\ell = 2$ must be independently measured before the trajectory of test bodies around stars can be predicted to high accuracy. 
For example, the NASA mission GRACE has measured the first 18 multipole moments of Earth~\cite{2004GeoRL..31.9607T}, which allows the modeling of satellite trajectories with errors of ${\cal{O}}(R_{\rm{E}}/r)^{19}$, with $r$ the distance from the satellite to Earth and $R_{\rm{E}}$ Earth's radius.  

\emph{The Hairs of Compact Stars.}~The non-applicability of the no-hair theorem extends, in principle, to all stars, including compact ones, such as neutron stars and strange quark stars. There is therefore no reason to expect stars to be bald, i.e.~for their exterior gravitational field to be independent of their internal structure or of their \emph{equations of state}, relating internal pressure and density. 

Recently, however, axisymmetric compact stars with isotropic pressure were shown to satisfy certain \emph{approximate} no-hair relations~\cite{I-Love-Q-Science,I-Love-Q-PRD,Pappas:2013naa,Stein:2014wpa,Yagi:2014bxa}. That is, all multipole moments in the multipole expansion of the exterior gravitational field of compact stars can be approximately expressed in terms of only three observable quantities. The first two are the mass 
monopole and current dipole of the compact star, while the third is the star's mass quadrupole moment. The independent measurement of these three observables allows us to determine all others through~\cite{Stein:2014wpa,Yagi:2014bxa} 
\be
\label{eq:NS-no-hair}
M_{\ell} + i \frac{\bar q}{a} S_{\ell} = B_{\lfloor \frac{\ell-1}{2} \rfloor} \; M \; \left(i \bar q \right)^{\ell}\,,
\ee
for $\ell>2$ and where $\lfloor x \rfloor$ stands for the largest integer not exceeding $x$, 
$\bar q \equiv - i (M_{2}/M)^{1/2}$, and $B_{\ell}$ is a pure number that in principle depends on the compact stars's equation of state. For example, Eq.~\eqref{eq:NS-no-hair} implies that $S_{3} = B_{1} S \bar q^{2}$, $M_{4} = B_{1} M \bar q^{4}$, etc. Unlike in the black hole case, $\bar q$ is independent of $a$.
 
These inter-relations between multipole moments correspond to compact star no-hair relations \emph{if and only if} they are independent of the compact star's internal structure. Equation~\eqref{eq:NS-no-hair} depends on the equation of state only through the coefficients $B_{\ell}$. These coefficients have been shown to take on \emph{approximately the same numerical value} (up to ten percent differences)~\cite{Stein:2014wpa,Yagi:2014bxa} within a very large class of compact star equations of state~\cite{APR,SLy,LS,Shen1,Shen2,Wiringa:1988tp,Alford:2004pf}. Such equation-of-state independence has been verified in a variety of scenarios~\cite{lattimer-lim,Chatziioannou:2014tha,Chan:2014tva,I-Love-Q-B,Pappas:2013naa,Chakrabarti:2013tca}. 
Thus, we say that compact stars are \emph{approximately bald or follicly challenged}.
 
But what is the origin of these approximate no-hair relations? Recently, Ref.~\cite{Yagi:2014qua} proposed a phenomenological picture to explain the approximate no-hair relations: as some set of parameters (such as the compactness or temperature of the star) are tuned beyond a given threshold, an \emph{approximate symmetry emerges} that is not present in general. In particular, they found that as compactness is increased, the eccentricity of isodensity contours inside the star, which is a measure of the degree of ellipticity of such contours, becomes nearly constant throughout the star, leading to the \emph{emergence of isodensity self-similarity}. 

\emph{Are the black hole and compact star no-hair relations related?}~If one pushes the phenomenological picture of~\cite{Yagi:2014qua} to the extreme and ``flows'' toward the black hole region of phase space by increasing compactness to 1/2 (the compactness of a non-spinning, Schwarzschild black hole), one encounters a problem. Black holes have all of their mass concentrated at their singularity, and thus, there is no matter density elsewhere in their interior with which to construct self-similar isodensity contours. It is thus not obvious or clear that the approximate no-hair relations will approach the black hole ones \emph{continuously} as an unstable compact star collapses into a black hole.

This topic was recently tackled in~\cite{Glampedakis:2013jya} by considering neutron stars with anisotropic pressure, since the latter allows for stars with compactnesses close to those of black holes. Reference~\cite{Glampedakis:2013jya} calculated the axisymmetric deformations of such an anisotropic star due to stellar rotation to leading-order in an expansion in powers of compactness. Their results indicated that incompressible strongly-anisotropic stars become \emph{prolate} (i.e.~ellipsoids with the semi-major axis aligned with the axis of rotation). If this remained true in the high-compactness regime, it would be in contrast to perturbed black holes, which have been long shown to be \emph{oblate} when deformed~\cite{Poisson:2009di}. The shape of a star is controlled, among other things, by its quadrupole moment; thus, if a neutron star with compactnesses close to that of a black hole is prolate instead of oblate, its quadrupole moments may not approach that of a black hole as the neutron star collapses. The work of~\cite{Glampedakis:2013jya} then suggests that the approximate no-hair relations for compact stars may not approach the black hole ones as the compactness increases. 

We here extend~\cite{Glampedakis:2013jya} by numerically constructing slowly-rotating anisotropic compact star solutions in full General Relativity. We first investigate how the stellar shape of anisotropic stars changes as one increases the stellar compactness. We then look at the relations among multipole moments for such stars and study how they approach the black hole limit as one increases the compactness. We carry out analytic calculations and explicitly show that the current dipole moment \emph{reaches} the black hole limit \emph{quadratically} in compactness. We conclude by pointing out the possibility of linking the results presented here to phase transitions in condensed matter physics.

\section{Anisotropic Compact Stars}

Let us revisit anisotropic stars with high compactness but this time in full General Relativity, instead of in a leading-order expansion in compactness. We concentrate on slowly-rotating stars, following the Hartle-Thorne approach~\cite{hartle1967,Hartle:1968ht}. Slowly-rotating, anisotropic neutron star solutions have been constructed in~\cite{Bayin:1982vw,Silva:2014fca} to linear order in spin, i.e.~in the ratio of the spin angular momentum to its mass squared, and we here extend such a calculation to third order for the first time. We also assume the stars are neutral, so that electromagnetic fields can be neglected. These are suitable approximations for old compact stars. We model matter anisotropy through the stress-energy tensor~\cite{Doneva:2012rd,Silva:2014fca}
\be
\label{eq:matter}
T_{\mu \nu} = \rho \; u_\mu u_\nu + p \; k_\mu k_\nu + q \; \Pi_{\mu \nu}\,,
\ee
where $\rho$ is the matter density, $p$ is the radial pressure (assumed to be barotropic $p=p(\rho)$), $q$ is the tangential pressure (responsible for anisotropy), $u^\mu$ is the fluid's four-velocity, $k^{\mu}$ is a radial vector, and $\Pi_{\mu \nu} = g_{\mu \nu} + u_\mu u_\nu - k_\mu k_\nu$ is a projection operator onto a two-dimensional surface orthogonal to $u^\mu$ and $k^\mu$. The unit radial vector is spacelike ($g^{\mu \nu} k_{\mu} k_\nu = 1$), while the four-velocity is timelike ($g^{\mu \nu} u_{\mu} u_{\nu} = -1$) and parameterized through $u^{\mu} = u^0 (1, 0,0,\Omega)$, where $u^{0}$ is a normalization constant and $\Omega$ is the spin angular velocity. 

Matter anisotropy is encoded in the tangential pressure, i.e.~that in the polar and azimuthal directions. Let us introduce the \emph{anisotropy parameter} $\sigma = p - q$, which we expand in the slow-rotation approximation.
We parameterize this function in the Bowers and Liang (BL)~\cite{1974ApJ...188..657B} framework
\be
\sigma_{\BL} =  \frac{\lambda_\BL}{3} \frac{(\rho + 3p) (\rho + p)}{1-{2M(r)}/{r}} r^2 + \Omega^2 f(h_{\mu \nu}) + {\cal{O}}(\Omega^4)\,,
\ee
where $\lambda_\BL$ is a constant that quantifies the amount of anisotropy, $M(r)$ is the mass interior to radius $r$ and $f(h_{\mu \nu})$ is a function of the metric perturbation at quadratic order in spin. One recovers isotropic stars when $\lambda_{\BL} = 0$, i.e. $f(h_{\mu \nu})$ vanishes when $\lambda_\BL = 0$, which forces $\sigma = 0$ and thus $q = p$, so that $T_{\mu \nu}$ is the stress-energy of a perfect fluid with isotropic pressure.

We here consider anisotropic compact stars because anisotropy allows us to explore the properties of stars with compactnesses close to those of black holes. In the BL model, incompressible stars reach the black hole limit when $\lambda_\BL = -2\pi$.  Reference~\cite{Glampedakis:2013jya} used this model in a leading-order expansion in compactness and found that the quadrupole moment changes sign at $\lambda_\BL = -0.8\pi$. Astrophysically, however, old compact stars are expected to be close to isotropic, with small anisotropy perhaps induced by stellar solid cores~\cite{1990sse..book.....K}, magnetic fields~\cite{Yazadjiev:2011ks}, phase transitions~\cite{Sawyer:1972cq,Carter:1998rn}, or two-fluid models (normal and superfluid components)~\cite{1980PhRvD..22..807L}.

With these models for matter anisotropy, we can now construct anisotropic compact star solutions in General Relativity. We first expand the Einstein field equations in the slow-rotation approximation following Hartle and Thorne~\cite{hartle1967,Hartle:1968ht}. We then specify a particular equation of state for the radial pressure, such as a tabulated one~\cite{APR,SLy,LS,Shen1,Shen2,Wiringa:1988tp,Alford:2004pf} or a polytropic one of the form $p = K \rho^{1+1/n}$, where $K$ and $n$ are the polytropic constant and index respectively. With this, we can now solve the expanded equations numerically and order by order in $\Omega$, adapting our numerical code developed in~\cite{I-Love-Q-Science,I-Love-Q-PRD,Yagi:2014bxa}. 
The numerical solution can then be used to extract the multipole moments of the compact star's exterior gravitational field far from the star~\cite{thorne-MM,Gursel,pappas-apostolatos}.
The details of the formalism can be found in~\cite{Yagi:2015hda}.

\begin{figure}[t]
\includegraphics[width=8.5cm,clip=true]{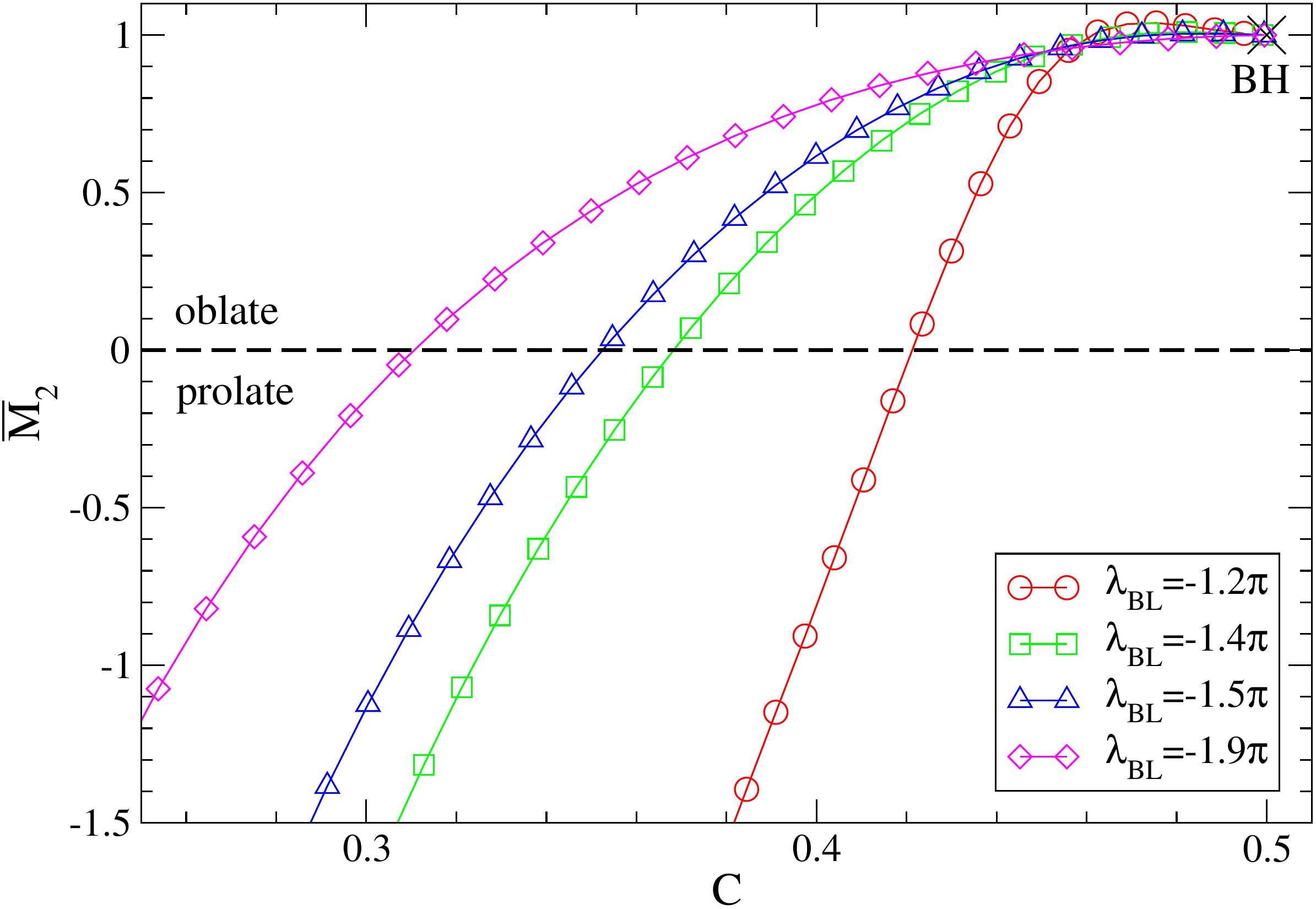}
\caption{\label{fig:C-dep-polyn0} (Color online) Compactness dependence of $\bar{M}_{2}$ for incompressible strongly anisotropic compact stars. The black hole value of $\bar{M}_{2}$ is shown with a cross on the top right corner. Although the compactness of anisotropic stars with $\lambda_\BL$ chosen here does not actually reach the black hole limit, as one would find by zooming to the $C \sim 0.5$ region, it does get very close to it.}
\end{figure}
\begin{figure*}[t]
\begin{center}
\includegraphics[width=8.5cm,clip=true]{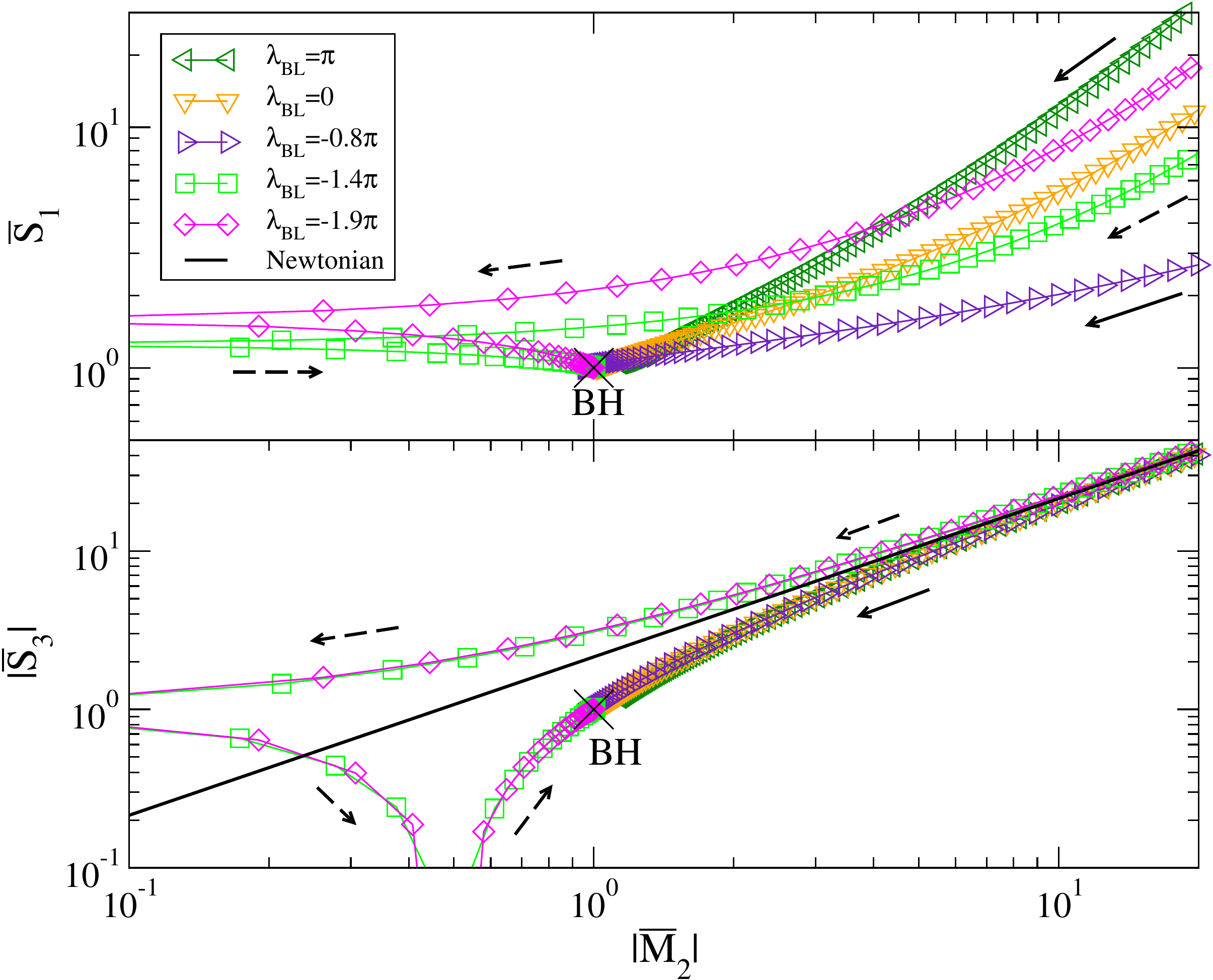}  
\includegraphics[width=8.8cm,clip=true]{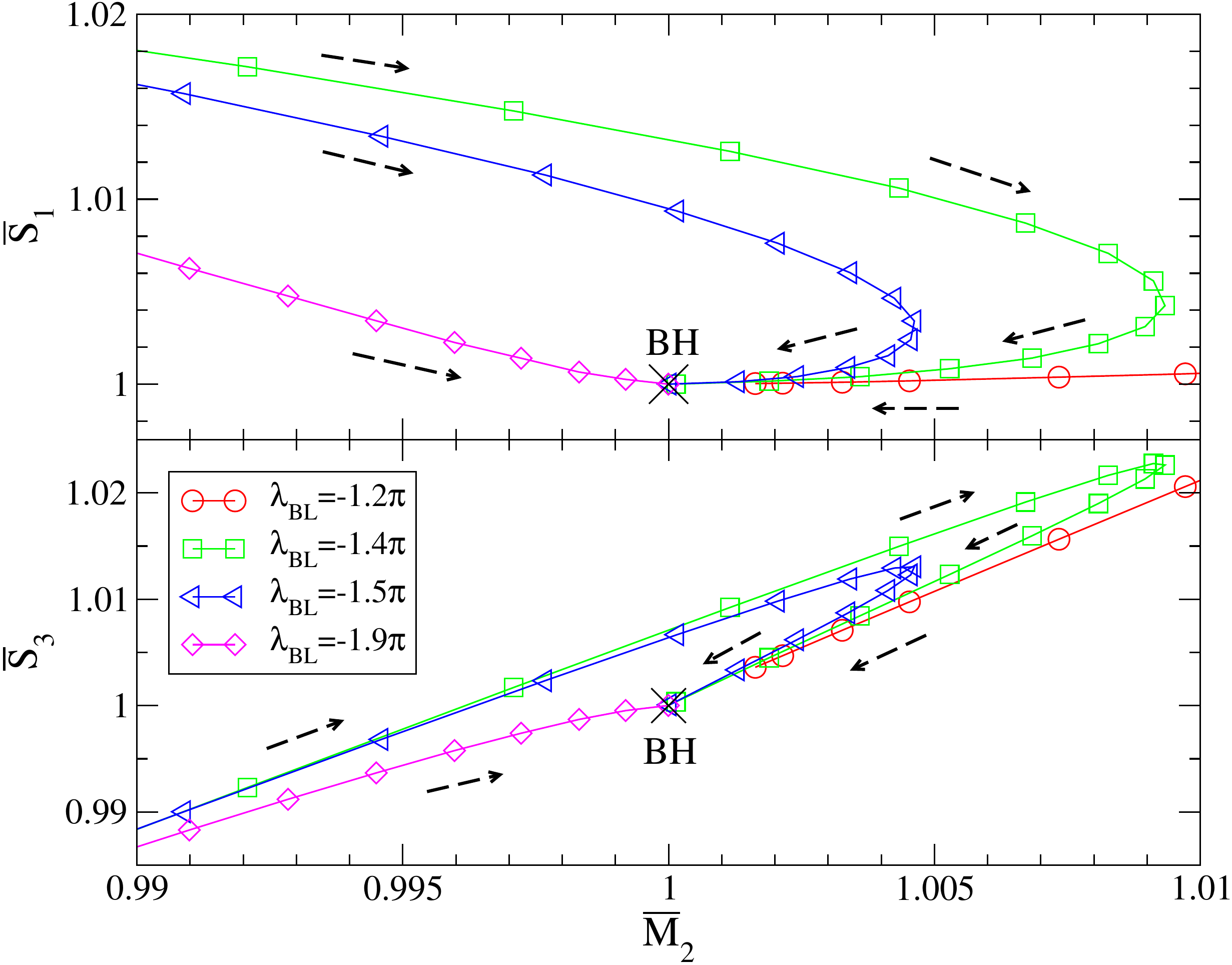}  
\caption{\label{fig:univ-polyn0} (Color online) ${\bar{S}}_{1}$-$\bar{M}_{2}$ (top left) and ${\bar{S}}_{3}$-$\bar{M}_{2}$ (bottom left) relations for incompressible anisotropic stars. The right panel zooms in to the black hole limit, shown with crosses. The arrows indicate the direction of increasing compactness, with the solid ones corresponding to stars with $\lambda_{\BL} \geq -0.8\pi $ and the dashed one for stars with $\lambda_{\BL} < -0.8\pi$. The solid line in the bottom left panel represents the analytic Newtonian relation, which is independent of $\lambda_{\BL}$. }
\end{center}
\end{figure*}

\section{Change in the stellar shape}

Figure~\ref{fig:C-dep-polyn0} shows the quadrupole moment normalized to its black hole value, $\bar{M}_{2} \equiv M_{2}/M_{2}^{\BH}$, as a function of stellar compactness, $C = M_{\NS}/R_{\NS}$, where $M_{\NS}$ and $R_{\NS}$ are the mass and radius of the compact star in the non-spinning configuration. Every point in this figure corresponds to the numerical construction of a compact star, with different compactness obtained by increasing the central density. Observe that $\bar{M}_{2}<0$ in the low compactness region when $\lambda_\BL < -0.8 \pi$, and thus the star is prolate as predicted by~\cite{Glampedakis:2013jya}. As compactness is increased, however, $\bar{M}_{2}$ becomes positive and the star becomes oblate. These results imply that relativistic corrections (proportional to high powers of compactness) change the results of~\cite{Glampedakis:2013jya}, forcing stars to become oblate and approach the black hole expectation.

Figure~\ref{fig:C-dep-polyn0} also shows that as the stellar compactness approaches $C_{\BH}$, the mass quadrupole approaches the black hole result, irrespective of the value of $\lambda_{\BL}$. We define the scaling exponent 
\be
k_{\bar{A}_{\ell}} = \lim_{\tau \to 0} \frac{\ln [\bar{A}_{\ell}(\tau)-1]}{\ln(\tau)}\,, \quad \tau \equiv \frac{C-C_\BH}{C_\BH}\,, 
\ee
in analogy with the critical exponents of second order phase transitions, where $\bar{A}_{\ell}$ is any of the mass multipoles $M_{\ell}$ or current multipoles $S_{\ell}$, normalized to their black hole values and $C_\BH = 1/2$ is the compactness of a non-rotating black hole. For example, $\bar{S}_{1} = S_{1}/S_{1}^{\BH}$, where $S_{1}^{\BH} = I^{\BH} \Omega = 4 M^{3} \Omega$~\cite{I-Love-Q-PRD} and $I^{\BH}$ is the moment of inertia of a non-rotating black hole, while  $\bar{S}_{3} = S_{3}/S_{3}^{\BH}$, where $S_{3}^{\BH} = - S^{3}/M^{2}$ by Eq.~\eqref{eq:no-hair}. Doing so, we find that $(\bar{A}_{\ell} -1 ) \propto \tau^{k_{\bar{A}{}_{\ell}}}$ near $\tau = 0$, with $k_{\bar{A}{}_{\ell}}$ presented in Table~\ref{table:exponent}. Observe that \emph{all} multipole moments of isotropic compact stars approach the critical point at the same rate -- as a fourth-order polynomial\footnote{Although the compactness of isotropic stars cannot \emph{reach} the black hole limit, $\bar{S}_1$ and $\bar{M}_2$ for such stars do \emph{approach} the limit, as can be seen from e.g. the bottom left panel of Fig.~9 in~\cite{I-Love-Q-PRD}.}. This behavior is approximately equation-of-state universal, with variations in $k_{\bar{A}_{\ell}}$ of only $\sim 10 \%$ due to the equation of state. Table~\ref{table:exponent} also shows that the mass and current multipole moments of strongly-anisotropic compact stars approach the black hole limit linearly and quadratically as one decreases the anisotropy parameter (corresponding to increasing the maximum compactness). 

{\renewcommand{\arraystretch}{1.2}
\begin{table}[b]
\begin{centering}
\begin{tabular}{c|c|cccc}
\hline
\hline
\noalign{\smallskip}
 &  \multicolumn{1}{c|}{Isotropic} &  \multicolumn{4}{c}{Anisotropic (poly. n=0)}   \\
$\lambda_\BL$ & 0 &  \multicolumn{1}{c}{$-1.2\pi$} &  \multicolumn{1}{c}{$-1.4\pi$}
&  \multicolumn{1}{c}{$-1.5\pi$} &  \multicolumn{1}{c}{$-1.9\pi$}   \\
\hline
$k_{\bar{S}_1}$ & $3.90 (\pm 0.49)$ & 2.20 & 2.08 & 2.03  & 1.87 \\
$k_{\bar{M}_2}$ & $4.22 (\pm 0.45)$ &1.32 & 1.12  & 1.08 & 1.15 \\
$k_{\bar{S}_3}$ & $4.19 (\pm 0.49)$ & 1.30 & 1.10  & 1.06 & 1.93   \\
\noalign{\smallskip}
\hline
\hline
\end{tabular}
\end{centering}
\caption{Scaling exponents of the multipole moments for compact stars. For isotropic stars, we present the averaged scaling exponent over various equations of state, with the maximum deviations from the mean denoted in parenthesis. For incompressible anisotropic stars, we present each scaling exponent as a function of the anisotropy parameter $\lambda_\BL$.}
\label{table:exponent}
\end{table}
}

In order to confirm whether the scaling exponent is an integer in the black hole limit, we analytically solved the Einstein equations to linear order in spin when $\lambda_\BL = -2\pi $, where the maximum compactness reaches the black hole compactness. The solution to such equations can be obtained in terms of hypergeometric functions. The reduced dipole moment $\bar S_1$ can then be written as a function of the stellar compactness as~\cite{in-prep}
\bw
\be
\label{eq:crit-exp-I}
\bar S_1 (C)  =  \frac{18\,{\it C}\,
{\mbox{$_2$F$_1$}(\frac{5}{2},{\frac {13}{4}};\,\frac{7}{2};\,2\,{\it C})}-9\,
{\mbox{$_2$F$_1$}(\frac{5}{2},{\frac {13}{4}};\,\frac{7}{2};\,2\,{\it C})}+5\,
{\mbox{$_2$F$_1$}(\frac{3}{2},\frac{9}{4};\,\frac{5}{2};\,2\,{\it C})} } {  8 C^2 \left[ 18\,{{
\it C}}^{2}
{\mbox{$_2$F$_1$}(\frac{5}{2},{\frac {13}{4}};\,\frac{7}{2};\,2\,{\it C})}-9\,{\it 
C}\,{\mbox{$_2$F$_1$}(\frac{5}{2},{\frac {13}{4}};\,\frac{7}{2};\,2\,{\it C})}+15\,{
\it C}\,{\mbox{$_2$F$_1$}(\frac{3}{2},\frac{9}{4};\,\frac{5}{2};\,2\,{\it C})}-5\,
{\mbox{$_2$F$_1$}(\frac{3}{2},\frac{9}{4};\,\frac{5}{2};\,2\,{\it C})} \right] }\,. 
\ee
\ew
From this equation, one can calculate the scaling exponent and exactly find $k_{\bar{S}_1} = 2$. This confirms that the scaling exponent for $\bar{S}_1$ reaches an integer.
We could not solve the Einstein equations at higher order in spin analytically to
derive similar expressions for $\bar M_2$ nor $\bar S_3$ since such higher order contributions are sourced by the complicated first order in spin solution presented above. 
These analytic results will be presented in more detail in~\cite{in-prep}.

\section{Approaching the Black Hole Limit}

Let us now investigate how the approximate no-hair relations for compact stars approach the black hole no-hair relations. To see this, let us explore a sequence of compact stars of increasing compactness and plot the first few low-$\ell$ multipole moments, normalized to the black hole values. 

Figure~\ref{fig:univ-polyn0} shows this sequence for incompressible anisotropic stars. As in Fig.~\ref{fig:C-dep-polyn0}, every point represents a different compact star constructed numerically. 
The arrows indicate the direction of increasing compactness, and the right panel is a zoom into the black hole limit region.  Observe that in the lower-compactness region (top right region of the bottom left panel) the multipole moments approach the approximate no-hair relations (shown as a black solid line labeled) expanded to lowest order in compactness, i.e.~the so-called \emph{Newtonian} approximate no-hair relations of~\cite{Stein:2014wpa} extended to anisotropic Newtonian stars. Observe also that isotropic stars tend to the black hole result, but do not reach it, since these stars do not reach sufficiently high compactnesses,  e.g.~the maximum compactness of a uniform density, isotropic star is 0.444~\cite{I-Love-Q-PRD}. 

Finally, observe that anisotropic stars, which can sample compactnesses much closer to $C_{\BH}$, continuously approach the black hole limit.  Interestingly, anisotropic stars with $\lambda_{\BL} \geq -0.8\pi$ approach the black hole limit directly (solid arrows), while those with $\lambda_{\BL} < -0.8\pi$ (dashed arrows) over shoot it, but then turn around and approach it. Recall that $\lambda_{\BL} = - 0.8 \pi$ was exactly the value of the anisotropic parameter at which deformed neutron stars switched from prolate to oblate in the low compactness region~\cite{Glampedakis:2013jya}. This means that low-compactness stars that are initially oblate ($\lambda_{\BL} > -0.8 \pi$) approach the limit directly, but those that are initially prolate ($\lambda_{\BL} < -0.8 \pi$) overshoot it and then approach it, while simultaneously becoming oblate. One can show analytically from Eq.~\eqref{eq:crit-exp-I} that $\bar S_1(1/2) = 1$, namely the current dipole moment reaches the black hole limit exactly as $C \to C_\BH$.

\section{Discussions and Future Directions}

The universal behavior in the scaling exponent that we found is analogous to the universal behavior observed in second-order phase transitions when considering critical phenomena, such as ferromagnetism. One may thus be able to understand the transition from compact stars to black holes as a phase transition, with $C$ and $C_\BH$ playing the role of temperature and critical temperature respectively. For example, Refs.~\cite{deBoer:2009wk,Arsiwalla:2010bt} studied a possible link between stellar transitions and second-order quantum phase transitions from a high density baryonic state into a thermal quark-gluon plasma state within the context of the AdS/CFT correspondence. We plan to investigate this further in future. 

Other future work could focus on constructing anisotropic compact star solutions to higher order in spin and see if higher multipole moments also approach the black hole limit as one increases the compactness. Another interesting avenue is to
show how the approximate no-hair relations for unstable compact stars change as they undergo gravitational collapse and become black holes \emph{dynamically}. Other future work includes whether this limiting behavior observed for anisotropic compact stars in General Relativity persists in other modified gravity theories. 

\acknowledgments
We would like to thank Neil Cornish, Bennet Link, Dana Longcope and Leo Stein for useful discussions. 
NY acknowledges support from NSF CAREER Award PHY-1250636.

\bibliography{master}

\begin{thebibliography}{50}%
\makeatletter
\providecommand \@ifxundefined [1]{%
 \@ifx{#1\undefined}
}%
\providecommand \@ifnum [1]{%
 \ifnum #1\expandafter \@firstoftwo
 \else \expandafter \@secondoftwo
 \fi
}%
\providecommand \@ifx [1]{%
 \ifx #1\expandafter \@firstoftwo
 \else \expandafter \@secondoftwo
 \fi
}%
\providecommand \natexlab [1]{#1}%
\providecommand \enquote  [1]{``#1''}%
\providecommand \bibnamefont  [1]{#1}%
\providecommand \bibfnamefont [1]{#1}%
\providecommand \citenamefont [1]{#1}%
\providecommand \href@noop [0]{\@secondoftwo}%
\providecommand \href [0]{\begingroup \@sanitize@url \@href}%
\providecommand \@href[1]{\@@startlink{#1}\@@href}%
\providecommand \@@href[1]{\endgroup#1\@@endlink}%
\providecommand \@sanitize@url [0]{\catcode `\\12\catcode `\$12\catcode
  `\&12\catcode `\#12\catcode `\^12\catcode `\_12\catcode `\%12\relax}%
\providecommand \@@startlink[1]{}%
\providecommand \@@endlink[0]{}%
\providecommand \url  [0]{\begingroup\@sanitize@url \@url }%
\providecommand \@url [1]{\endgroup\@href {#1}{\urlprefix }}%
\providecommand \urlprefix  [0]{URL }%
\providecommand \Eprint [0]{\href }%
\providecommand \doibase [0]{http://dx.doi.org/}%
\providecommand \selectlanguage [0]{\@gobble}%
\providecommand \bibinfo  [0]{\@secondoftwo}%
\providecommand \bibfield  [0]{\@secondoftwo}%
\providecommand \translation [1]{[#1]}%
\providecommand \BibitemOpen [0]{}%
\providecommand \bibitemStop [0]{}%
\providecommand \bibitemNoStop [0]{.\EOS\space}%
\providecommand \EOS [0]{\spacefactor3000\relax}%
\providecommand \BibitemShut  [1]{\csname bibitem#1\endcsname}%
\let\auto@bib@innerbib\@empty
\bibitem [{\citenamefont {Will}(2006)}]{will-living}%
  \BibitemOpen
  \bibfield  {author} {\bibinfo {author} {\bibfnamefont {C.~M.}\ \bibnamefont
  {Will}},\ }\href {http://www.livingreviews.org/lrr-2006-3} {\bibfield
  {journal} {\bibinfo  {journal} {Living Reviews in Relativity}\ }\textbf
  {\bibinfo {volume} {9}},\ \bibinfo {pages} {3} (\bibinfo {year} {2006})},\
  \Eprint {http://arxiv.org/abs/gr-qc/0510072} {arXiv:gr-qc/0510072}
  \BibitemShut {NoStop}%
\bibitem [{\citenamefont {Liu}\ \emph {et~al.}(2012)\citenamefont {Liu},
  \citenamefont {Wex}, \citenamefont {Kramer}, \citenamefont {Cordes},\ and\
  \citenamefont {Lazio}}]{Liu:2011ae}%
  \BibitemOpen
  \bibfield  {author} {\bibinfo {author} {\bibfnamefont {K.}~\bibnamefont
  {Liu}}, \bibinfo {author} {\bibfnamefont {N.}~\bibnamefont {Wex}}, \bibinfo
  {author} {\bibfnamefont {M.}~\bibnamefont {Kramer}}, \bibinfo {author}
  {\bibfnamefont {J.}~\bibnamefont {Cordes}}, \ and\ \bibinfo {author}
  {\bibfnamefont {T.}~\bibnamefont {Lazio}},\ }\href {\doibase
  10.1088/0004-637X/747/1/1} {\bibfield  {journal} {\bibinfo  {journal}
  {Astrophys.J.}\ }\textbf {\bibinfo {volume} {747}},\ \bibinfo {pages} {1}
  (\bibinfo {year} {2012})},\ \Eprint {http://arxiv.org/abs/1112.2151}
  {arXiv:1112.2151 [astro-ph.HE]} \BibitemShut {NoStop}%
\bibitem [{\citenamefont {Will}(2008)}]{Will:2007pp}%
  \BibitemOpen
  \bibfield  {author} {\bibinfo {author} {\bibfnamefont {C.~M.}\ \bibnamefont
  {Will}},\ }\href {\doibase 10.1086/528847} {\bibfield  {journal} {\bibinfo
  {journal} {Astrophys.J.}\ }\textbf {\bibinfo {volume} {674}},\ \bibinfo
  {pages} {L25} (\bibinfo {year} {2008})},\ \Eprint
  {http://arxiv.org/abs/0711.1677} {arXiv:0711.1677 [astro-ph]} \BibitemShut
  {NoStop}%
\bibitem [{\citenamefont {Misner}\ \emph {et~al.}(1973)\citenamefont {Misner},
  \citenamefont {Thorne},\ and\ \citenamefont {Wheeler}}]{MTW}%
  \BibitemOpen
  \bibfield  {author} {\bibinfo {author} {\bibfnamefont {C.~W.}\ \bibnamefont
  {Misner}}, \bibinfo {author} {\bibfnamefont {K.}~\bibnamefont {Thorne}}, \
  and\ \bibinfo {author} {\bibfnamefont {J.~A.}\ \bibnamefont {Wheeler}},\
  }\href@noop {} {\emph {\bibinfo {title} {Gravitation}}}\ (\bibinfo
  {publisher} {W. H. Freeman \& Co.},\ \bibinfo {address} {San Francisco},\
  \bibinfo {year} {1973})\BibitemShut {NoStop}%
\bibitem [{\citenamefont {Carter}(1971)}]{carter-uniqueness}%
  \BibitemOpen
  \bibfield  {author} {\bibinfo {author} {\bibfnamefont {B.}~\bibnamefont
  {Carter}},\ }\href {\doibase 10.1103/PhysRevLett.26.331} {\bibfield
  {journal} {\bibinfo  {journal} {Phys.Rev.Lett.}\ }\textbf {\bibinfo {volume}
  {26}},\ \bibinfo {pages} {331} (\bibinfo {year} {1971})}\BibitemShut
  {NoStop}%
\bibitem [{\citenamefont {Hansen}(1974)}]{hansen}%
  \BibitemOpen
  \bibfield  {author} {\bibinfo {author} {\bibfnamefont {R.~O.}\ \bibnamefont
  {Hansen}},\ }\href@noop {} {\bibfield  {journal} {\bibinfo  {journal} {J.
  Math. Phys.}\ }\textbf {\bibinfo {volume} {15}},\ \bibinfo {pages} {46}
  (\bibinfo {year} {1974})}\BibitemShut {NoStop}%
\bibitem [{\citenamefont {Robinson}(1975)}]{robinson}%
  \BibitemOpen
  \bibfield  {author} {\bibinfo {author} {\bibfnamefont {D.}~\bibnamefont
  {Robinson}},\ }\href {\doibase 10.1103/PhysRevLett.34.905} {\bibfield
  {journal} {\bibinfo  {journal} {Phys.Rev.Lett.}\ }\textbf {\bibinfo {volume}
  {34}},\ \bibinfo {pages} {905} (\bibinfo {year} {1975})}\BibitemShut
  {NoStop}%
\bibitem [{\citenamefont {Israel}(1967)}]{israel}%
  \BibitemOpen
  \bibfield  {author} {\bibinfo {author} {\bibfnamefont {W.}~\bibnamefont
  {Israel}},\ }\href {\doibase 10.1103/PhysRev.164.1776} {\bibfield  {journal}
  {\bibinfo  {journal} {Phys. Rev.}\ }\textbf {\bibinfo {volume} {164}},\
  \bibinfo {pages} {1776} (\bibinfo {year} {1967})}\BibitemShut {NoStop}%
\bibitem [{\citenamefont {Israel}(1968)}]{israel2}%
  \BibitemOpen
  \bibfield  {author} {\bibinfo {author} {\bibfnamefont {W.}~\bibnamefont
  {Israel}},\ }\href@noop {} {\bibfield  {journal} {\bibinfo  {journal}
  {Commun.Math.Phys.}\ }\textbf {\bibinfo {volume} {8}},\ \bibinfo {pages}
  {245} (\bibinfo {year} {1968})}\BibitemShut {NoStop}%
\bibitem [{\citenamefont {Hawking}(1971)}]{hawking-uniqueness0}%
  \BibitemOpen
  \bibfield  {author} {\bibinfo {author} {\bibfnamefont {S.}~\bibnamefont
  {Hawking}},\ }\href {\doibase 10.1103/PhysRevLett.26.1344} {\bibfield
  {journal} {\bibinfo  {journal} {Phys.Rev.Lett.}\ }\textbf {\bibinfo {volume}
  {26}},\ \bibinfo {pages} {1344} (\bibinfo {year} {1971})}\BibitemShut
  {NoStop}%
\bibitem [{\citenamefont {Hawking}(1972)}]{hawking-uniqueness}%
  \BibitemOpen
  \bibfield  {author} {\bibinfo {author} {\bibfnamefont {S.~W.}\ \bibnamefont
  {Hawking}},\ }\href {\doibase 10.1007/BF01877517} {\bibfield  {journal}
  {\bibinfo  {journal} {Commun. Math. Phys.}\ }\textbf {\bibinfo {volume}
  {25}},\ \bibinfo {pages} {152} (\bibinfo {year} {1972})}\BibitemShut
  {NoStop}%
\bibitem [{\citenamefont {{Tapley}}\ \emph {et~al.}(2004)\citenamefont
  {{Tapley}}, \citenamefont {{Bettadpur}}, \citenamefont {{Watkins}},\ and\
  \citenamefont {{Reigber}}}]{2004GeoRL..31.9607T}%
  \BibitemOpen
  \bibfield  {author} {\bibinfo {author} {\bibfnamefont {B.~D.}\ \bibnamefont
  {{Tapley}}}, \bibinfo {author} {\bibfnamefont {S.}~\bibnamefont
  {{Bettadpur}}}, \bibinfo {author} {\bibfnamefont {M.}~\bibnamefont
  {{Watkins}}}, \ and\ \bibinfo {author} {\bibfnamefont {C.}~\bibnamefont
  {{Reigber}}},\ }\href {\doibase 10.1029/2004GL019920} {\bibfield  {journal}
  {\bibinfo  {journal} {Geophysical Research Letters}\ }\textbf {\bibinfo
  {volume} {31}},\ \bibinfo {eid} {L09607} (\bibinfo {year}
  {2004})}\BibitemShut {NoStop}%
\bibitem [{\citenamefont {Yagi}\ and\ \citenamefont
  {Yunes}(2013)}]{I-Love-Q-Science}%
  \BibitemOpen
  \bibfield  {author} {\bibinfo {author} {\bibfnamefont {K.}~\bibnamefont
  {Yagi}}\ and\ \bibinfo {author} {\bibfnamefont {N.}~\bibnamefont {Yunes}},\
  }\href {\doibase 10.1126/science.1236462} {\bibfield  {journal} {\bibinfo
  {journal} {Science}\ }\textbf {\bibinfo {volume} {341}},\ \bibinfo {pages}
  {365} (\bibinfo {year} {2013})},\ \Eprint {http://arxiv.org/abs/1302.4499}
  {arXiv:1302.4499 [gr-qc]} \BibitemShut {NoStop}%
\bibitem [{\citenamefont {{Yagi}}\ and\ \citenamefont
  {{Yunes}}(2013)}]{I-Love-Q-PRD}%
  \BibitemOpen
  \bibfield  {author} {\bibinfo {author} {\bibfnamefont {K.}~\bibnamefont
  {{Yagi}}}\ and\ \bibinfo {author} {\bibfnamefont {N.}~\bibnamefont
  {{Yunes}}},\ }\href {\doibase 10.1103/PhysRevD.88.023009} {\bibfield
  {journal} {\bibinfo  {journal} {\prd}\ }\textbf {\bibinfo {volume} {88}},\
  \bibinfo {eid} {023009} (\bibinfo {year} {2013})},\ \Eprint
  {http://arxiv.org/abs/1303.1528} {arXiv:1303.1528 [gr-qc]} \BibitemShut
  {NoStop}%
\bibitem [{\citenamefont {Pappas}\ and\ \citenamefont
  {Apostolatos}(2014)}]{Pappas:2013naa}%
  \BibitemOpen
  \bibfield  {author} {\bibinfo {author} {\bibfnamefont {G.}~\bibnamefont
  {Pappas}}\ and\ \bibinfo {author} {\bibfnamefont {T.~A.}\ \bibnamefont
  {Apostolatos}},\ }\href {\doibase 10.1103/PhysRevLett.112.121101} {\bibfield
  {journal} {\bibinfo  {journal} {Phys.Rev.Lett.}\ }\textbf {\bibinfo {volume}
  {112}},\ \bibinfo {pages} {121101} (\bibinfo {year} {2014})},\ \Eprint
  {http://arxiv.org/abs/1311.5508} {arXiv:1311.5508 [gr-qc]} \BibitemShut
  {NoStop}%
\bibitem [{\citenamefont {Stein}\ \emph {et~al.}(2014)\citenamefont {Stein},
  \citenamefont {Yagi},\ and\ \citenamefont {Yunes}}]{Stein:2014wpa}%
  \BibitemOpen
  \bibfield  {author} {\bibinfo {author} {\bibfnamefont {L.~C.}\ \bibnamefont
  {Stein}}, \bibinfo {author} {\bibfnamefont {K.}~\bibnamefont {Yagi}}, \ and\
  \bibinfo {author} {\bibfnamefont {N.}~\bibnamefont {Yunes}},\ }\href
  {\doibase 10.1088/0004-637X/788/1/15} {\bibfield  {journal} {\bibinfo
  {journal} {Astrophys.J.}\ }\textbf {\bibinfo {volume} {788}},\ \bibinfo
  {pages} {15} (\bibinfo {year} {2014})},\ \Eprint
  {http://arxiv.org/abs/1312.4532} {arXiv:1312.4532 [gr-qc]} \BibitemShut
  {NoStop}%
\bibitem [{\citenamefont {Yagi}\ \emph
  {et~al.}(2014{\natexlab{a}})\citenamefont {Yagi}, \citenamefont {Kyutoku},
  \citenamefont {Pappas}, \citenamefont {Yunes},\ and\ \citenamefont
  {Apostolatos}}]{Yagi:2014bxa}%
  \BibitemOpen
  \bibfield  {author} {\bibinfo {author} {\bibfnamefont {K.}~\bibnamefont
  {Yagi}}, \bibinfo {author} {\bibfnamefont {K.}~\bibnamefont {Kyutoku}},
  \bibinfo {author} {\bibfnamefont {G.}~\bibnamefont {Pappas}}, \bibinfo
  {author} {\bibfnamefont {N.}~\bibnamefont {Yunes}}, \ and\ \bibinfo {author}
  {\bibfnamefont {T.~A.}\ \bibnamefont {Apostolatos}},\ }\href {\doibase
  10.1103/PhysRevD.89.124013} {\bibfield  {journal} {\bibinfo  {journal}
  {Phys.Rev.}\ }\textbf {\bibinfo {volume} {D89}},\ \bibinfo {pages} {124013}
  (\bibinfo {year} {2014}{\natexlab{a}})},\ \Eprint
  {http://arxiv.org/abs/1403.6243} {arXiv:1403.6243 [gr-qc]} \BibitemShut
  {NoStop}%
\bibitem [{\citenamefont {Akmal}\ \emph {et~al.}(1998)\citenamefont {Akmal},
  \citenamefont {Pandharipande},\ and\ \citenamefont {Ravenhall}}]{APR}%
  \BibitemOpen
  \bibfield  {author} {\bibinfo {author} {\bibfnamefont {A.}~\bibnamefont
  {Akmal}}, \bibinfo {author} {\bibfnamefont {V.}~\bibnamefont
  {Pandharipande}}, \ and\ \bibinfo {author} {\bibfnamefont {D.}~\bibnamefont
  {Ravenhall}},\ }\href {\doibase 10.1103/PhysRevC.58.1804} {\bibfield
  {journal} {\bibinfo  {journal} {Phys.Rev.}\ }\textbf {\bibinfo {volume}
  {C58}},\ \bibinfo {pages} {1804} (\bibinfo {year} {1998})},\ \Eprint
  {http://arxiv.org/abs/nucl-th/9804027} {arXiv:nucl-th/9804027 [nucl-th]}
  \BibitemShut {NoStop}%
\bibitem [{\citenamefont {{Douchin}}\ and\ \citenamefont
  {{Haensel}}(2001)}]{SLy}%
  \BibitemOpen
  \bibfield  {author} {\bibinfo {author} {\bibfnamefont {F.}~\bibnamefont
  {{Douchin}}}\ and\ \bibinfo {author} {\bibfnamefont {P.}~\bibnamefont
  {{Haensel}}},\ }\href {\doibase 10.1051/0004-6361:20011402} {\bibfield
  {journal} {\bibinfo  {journal} {Astron. Astrophys.}\ }\textbf {\bibinfo
  {volume} {380}},\ \bibinfo {pages} {151} (\bibinfo {year}
  {2001})}\BibitemShut {NoStop}%
\bibitem [{\citenamefont {{Lattimer}}\ and\ \citenamefont {{Douglas
  Swesty}}(1991)}]{LS}%
  \BibitemOpen
  \bibfield  {author} {\bibinfo {author} {\bibfnamefont {J.~M.}\ \bibnamefont
  {{Lattimer}}}\ and\ \bibinfo {author} {\bibfnamefont {F.}~\bibnamefont
  {{Douglas Swesty}}},\ }\href {\doibase 10.1016/0375-9474(91)90452-C}
  {\bibfield  {journal} {\bibinfo  {journal} {Nuclear Physics A}\ }\textbf
  {\bibinfo {volume} {535}},\ \bibinfo {pages} {331} (\bibinfo {year}
  {1991})}\BibitemShut {NoStop}%
\bibitem [{\citenamefont {{Shen}}\ \emph
  {et~al.}(1998{\natexlab{a}})\citenamefont {{Shen}}, \citenamefont {{Toki}},
  \citenamefont {{Oyamatsu}},\ and\ \citenamefont {{Sumiyoshi}}}]{Shen1}%
  \BibitemOpen
  \bibfield  {author} {\bibinfo {author} {\bibfnamefont {H.}~\bibnamefont
  {{Shen}}}, \bibinfo {author} {\bibfnamefont {H.}~\bibnamefont {{Toki}}},
  \bibinfo {author} {\bibfnamefont {K.}~\bibnamefont {{Oyamatsu}}}, \ and\
  \bibinfo {author} {\bibfnamefont {K.}~\bibnamefont {{Sumiyoshi}}},\ }\href
  {\doibase 10.1016/S0375-9474(98)00236-X} {\bibfield  {journal} {\bibinfo
  {journal} {Nuclear Physics A}\ }\textbf {\bibinfo {volume} {637}},\ \bibinfo
  {pages} {435} (\bibinfo {year} {1998}{\natexlab{a}})}\BibitemShut {NoStop}%
\bibitem [{\citenamefont {{Shen}}\ \emph
  {et~al.}(1998{\natexlab{b}})\citenamefont {{Shen}}, \citenamefont {{Toki}},
  \citenamefont {{Oyamatsu}},\ and\ \citenamefont {{Sumiyoshi}}}]{Shen2}%
  \BibitemOpen
  \bibfield  {author} {\bibinfo {author} {\bibfnamefont {H.}~\bibnamefont
  {{Shen}}}, \bibinfo {author} {\bibfnamefont {H.}~\bibnamefont {{Toki}}},
  \bibinfo {author} {\bibfnamefont {K.}~\bibnamefont {{Oyamatsu}}}, \ and\
  \bibinfo {author} {\bibfnamefont {K.}~\bibnamefont {{Sumiyoshi}}},\ }\href
  {\doibase 10.1143/PTP.100.1013} {\bibfield  {journal} {\bibinfo  {journal}
  {Progress of Theoretical Physics}\ }\textbf {\bibinfo {volume} {100}},\
  \bibinfo {pages} {1013} (\bibinfo {year} {1998}{\natexlab{b}})}\BibitemShut
  {NoStop}%
\bibitem [{\citenamefont {Wiringa}\ \emph {et~al.}(1988)\citenamefont
  {Wiringa}, \citenamefont {Fiks},\ and\ \citenamefont
  {Fabrocini}}]{Wiringa:1988tp}%
  \BibitemOpen
  \bibfield  {author} {\bibinfo {author} {\bibfnamefont {R.~B.}\ \bibnamefont
  {Wiringa}}, \bibinfo {author} {\bibfnamefont {V.}~\bibnamefont {Fiks}}, \
  and\ \bibinfo {author} {\bibfnamefont {A.}~\bibnamefont {Fabrocini}},\ }\href
  {\doibase 10.1103/PhysRevC.38.1010} {\bibfield  {journal} {\bibinfo
  {journal} {Phys.Rev.}\ }\textbf {\bibinfo {volume} {C38}},\ \bibinfo {pages}
  {1010} (\bibinfo {year} {1988})}\BibitemShut {NoStop}%
\bibitem [{\citenamefont {Alford}\ \emph {et~al.}(2005)\citenamefont {Alford},
  \citenamefont {Braby}, \citenamefont {Paris},\ and\ \citenamefont
  {Reddy}}]{Alford:2004pf}%
  \BibitemOpen
  \bibfield  {author} {\bibinfo {author} {\bibfnamefont {M.}~\bibnamefont
  {Alford}}, \bibinfo {author} {\bibfnamefont {M.}~\bibnamefont {Braby}},
  \bibinfo {author} {\bibfnamefont {M.}~\bibnamefont {Paris}}, \ and\ \bibinfo
  {author} {\bibfnamefont {S.}~\bibnamefont {Reddy}},\ }\href {\doibase
  10.1086/430902} {\bibfield  {journal} {\bibinfo  {journal} {Astrophys.J.}\
  }\textbf {\bibinfo {volume} {629}},\ \bibinfo {pages} {969} (\bibinfo {year}
  {2005})},\ \Eprint {http://arxiv.org/abs/nucl-th/0411016}
  {arXiv:nucl-th/0411016 [nucl-th]} \BibitemShut {NoStop}%
\bibitem [{\citenamefont {Lattimer}\ and\ \citenamefont
  {Lim}(2013)}]{lattimer-lim}%
  \BibitemOpen
  \bibfield  {author} {\bibinfo {author} {\bibfnamefont {J.~M.}\ \bibnamefont
  {Lattimer}}\ and\ \bibinfo {author} {\bibfnamefont {Y.}~\bibnamefont {Lim}},\
  }\href {\doibase 10.1088/0004-637X/771/1/51} {\bibfield  {journal} {\bibinfo
  {journal} {ApJ. 771,}\ }\textbf {\bibinfo {volume} {51}} (\bibinfo {year}
  {2013}),\ 10.1088/0004-637X/771/1/51},\ \Eprint
  {http://arxiv.org/abs/1203.4286} {arXiv:1203.4286 [nucl-th]} \BibitemShut
  {NoStop}%
\bibitem [{\citenamefont {Chatziioannou}\ \emph {et~al.}(2014)\citenamefont
  {Chatziioannou}, \citenamefont {Yagi},\ and\ \citenamefont
  {Yunes}}]{Chatziioannou:2014tha}%
  \BibitemOpen
  \bibfield  {author} {\bibinfo {author} {\bibfnamefont {K.}~\bibnamefont
  {Chatziioannou}}, \bibinfo {author} {\bibfnamefont {K.}~\bibnamefont {Yagi}},
  \ and\ \bibinfo {author} {\bibfnamefont {N.}~\bibnamefont {Yunes}},\ }\href
  {\doibase 10.1103/PhysRevD.90.064030} {\bibfield  {journal} {\bibinfo
  {journal} {Phys.Rev.}\ }\textbf {\bibinfo {volume} {D90}},\ \bibinfo {pages}
  {064030} (\bibinfo {year} {2014})},\ \Eprint {http://arxiv.org/abs/1406.7135}
  {arXiv:1406.7135 [gr-qc]} \BibitemShut {NoStop}%
\bibitem [{\citenamefont {Chan}\ \emph {et~al.}(2015)\citenamefont {Chan},
  \citenamefont {Chan},\ and\ \citenamefont {Leung}}]{Chan:2014tva}%
  \BibitemOpen
  \bibfield  {author} {\bibinfo {author} {\bibfnamefont {T.}~\bibnamefont
  {Chan}}, \bibinfo {author} {\bibfnamefont {A.~P.}\ \bibnamefont {Chan}}, \
  and\ \bibinfo {author} {\bibfnamefont {P.}~\bibnamefont {Leung}},\ }\href
  {\doibase 10.1103/PhysRevD.91.044017} {\bibfield  {journal} {\bibinfo
  {journal} {Phys.Rev.}\ }\textbf {\bibinfo {volume} {D91}},\ \bibinfo {pages}
  {044017} (\bibinfo {year} {2015})},\ \Eprint {http://arxiv.org/abs/1411.7141}
  {arXiv:1411.7141 [astro-ph.SR]} \BibitemShut {NoStop}%
\bibitem [{\citenamefont {{Haskell}}\ \emph {et~al.}(2014)\citenamefont
  {{Haskell}}, \citenamefont {{Ciolfi}}, \citenamefont {{Pannarale}},\ and\
  \citenamefont {{Rezzolla}}}]{I-Love-Q-B}%
  \BibitemOpen
  \bibfield  {author} {\bibinfo {author} {\bibfnamefont {B.}~\bibnamefont
  {{Haskell}}}, \bibinfo {author} {\bibfnamefont {R.}~\bibnamefont {{Ciolfi}}},
  \bibinfo {author} {\bibfnamefont {F.}~\bibnamefont {{Pannarale}}}, \ and\
  \bibinfo {author} {\bibfnamefont {L.}~\bibnamefont {{Rezzolla}}},\ }\href
  {\doibase 10.1093/mnrasl/slt161} {\bibfield  {journal} {\bibinfo  {journal}
  {Mon. Not. Roy. Astron. Soc.}\ }\textbf {\bibinfo {volume} {438}},\ \bibinfo
  {pages} {L71} (\bibinfo {year} {2014})},\ \Eprint
  {http://arxiv.org/abs/1309.3885} {arXiv:1309.3885 [astro-ph.SR]} \BibitemShut
  {NoStop}%
\bibitem [{\citenamefont {Chakrabarti}\ \emph {et~al.}(2014)\citenamefont
  {Chakrabarti}, \citenamefont {Delsate}, \citenamefont {Gurlebeck},\ and\
  \citenamefont {Steinhoff}}]{Chakrabarti:2013tca}%
  \BibitemOpen
  \bibfield  {author} {\bibinfo {author} {\bibfnamefont {S.}~\bibnamefont
  {Chakrabarti}}, \bibinfo {author} {\bibfnamefont {T.}~\bibnamefont
  {Delsate}}, \bibinfo {author} {\bibfnamefont {N.}~\bibnamefont {Gurlebeck}},
  \ and\ \bibinfo {author} {\bibfnamefont {J.}~\bibnamefont {Steinhoff}},\
  }\href {\doibase 10.1103/PhysRevLett.112.201102} {\bibfield  {journal}
  {\bibinfo  {journal} {Phys.Rev.Lett.}\ }\textbf {\bibinfo {volume} {112}},\
  \bibinfo {pages} {201102} (\bibinfo {year} {2014})},\ \Eprint
  {http://arxiv.org/abs/1311.6509} {arXiv:1311.6509 [gr-qc]} \BibitemShut
  {NoStop}%
\bibitem [{\citenamefont {Yagi}\ \emph
  {et~al.}(2014{\natexlab{b}})\citenamefont {Yagi}, \citenamefont {Stein},
  \citenamefont {Pappas}, \citenamefont {Yunes},\ and\ \citenamefont
  {Apostolatos}}]{Yagi:2014qua}%
  \BibitemOpen
  \bibfield  {author} {\bibinfo {author} {\bibfnamefont {K.}~\bibnamefont
  {Yagi}}, \bibinfo {author} {\bibfnamefont {L.~C.}\ \bibnamefont {Stein}},
  \bibinfo {author} {\bibfnamefont {G.}~\bibnamefont {Pappas}}, \bibinfo
  {author} {\bibfnamefont {N.}~\bibnamefont {Yunes}}, \ and\ \bibinfo {author}
  {\bibfnamefont {T.~A.}\ \bibnamefont {Apostolatos}},\ }\href {\doibase
  10.1103/PhysRevD.90.063010} {\bibfield  {journal} {\bibinfo  {journal}
  {Phys.Rev.}\ }\textbf {\bibinfo {volume} {D90}},\ \bibinfo {pages} {063010}
  (\bibinfo {year} {2014}{\natexlab{b}})},\ \Eprint
  {http://arxiv.org/abs/1406.7587} {arXiv:1406.7587 [gr-qc]} \BibitemShut
  {NoStop}%
\bibitem [{\citenamefont {Glampedakis}\ \emph {et~al.}(2014)\citenamefont
  {Glampedakis}, \citenamefont {Kapadia},\ and\ \citenamefont
  {Kennefick}}]{Glampedakis:2013jya}%
  \BibitemOpen
  \bibfield  {author} {\bibinfo {author} {\bibfnamefont {K.}~\bibnamefont
  {Glampedakis}}, \bibinfo {author} {\bibfnamefont {S.~J.}\ \bibnamefont
  {Kapadia}}, \ and\ \bibinfo {author} {\bibfnamefont {D.}~\bibnamefont
  {Kennefick}},\ }\href {\doibase 10.1103/PhysRevD.89.024007} {\bibfield
  {journal} {\bibinfo  {journal} {Phys.Rev.}\ }\textbf {\bibinfo {volume}
  {D89}},\ \bibinfo {pages} {024007} (\bibinfo {year} {2014})},\ \Eprint
  {http://arxiv.org/abs/1312.1912} {arXiv:1312.1912 [gr-qc]} \BibitemShut
  {NoStop}%
\bibitem [{\citenamefont {Poisson}(2009)}]{Poisson:2009di}%
  \BibitemOpen
  \bibfield  {author} {\bibinfo {author} {\bibfnamefont {E.}~\bibnamefont
  {Poisson}},\ }\href {\doibase 10.1103/PhysRevD.80.064029} {\bibfield
  {journal} {\bibinfo  {journal} {Phys.Rev.}\ }\textbf {\bibinfo {volume}
  {D80}},\ \bibinfo {pages} {064029} (\bibinfo {year} {2009})},\ \Eprint
  {http://arxiv.org/abs/0907.0874} {arXiv:0907.0874 [gr-qc]} \BibitemShut
  {NoStop}%
\bibitem [{\citenamefont {Hartle}(1967)}]{hartle1967}%
  \BibitemOpen
  \bibfield  {author} {\bibinfo {author} {\bibfnamefont {J.~B.}\ \bibnamefont
  {Hartle}},\ }\href@noop {} {\bibfield  {journal} {\bibinfo  {journal}
  {Astrophys.J.}\ }\textbf {\bibinfo {volume} {150}},\ \bibinfo {pages} {1005}
  (\bibinfo {year} {1967})}\BibitemShut {NoStop}%
\bibitem [{\citenamefont {{Hartle}}\ and\ \citenamefont
  {{Thorne}}(1968)}]{Hartle:1968ht}%
  \BibitemOpen
  \bibfield  {author} {\bibinfo {author} {\bibfnamefont {J.~B.}\ \bibnamefont
  {{Hartle}}}\ and\ \bibinfo {author} {\bibfnamefont {K.~S.}\ \bibnamefont
  {{Thorne}}},\ }\href@noop {} {\bibfield  {journal} {\bibinfo  {journal}
  {\apj}\ }\textbf {\bibinfo {volume} {153}},\ \bibinfo {pages} {807} (\bibinfo
  {year} {1968})}\BibitemShut {NoStop}%
\bibitem [{\citenamefont {Bayin}(1982)}]{Bayin:1982vw}%
  \BibitemOpen
  \bibfield  {author} {\bibinfo {author} {\bibfnamefont {S.~S.}\ \bibnamefont
  {Bayin}},\ }\href {\doibase 10.1103/PhysRevD.26.1262} {\bibfield  {journal}
  {\bibinfo  {journal} {Phys.Rev.}\ }\textbf {\bibinfo {volume} {D26}},\
  \bibinfo {pages} {1262} (\bibinfo {year} {1982})}\BibitemShut {NoStop}%
\bibitem [{\citenamefont {Silva}\ \emph {et~al.}(2015)\citenamefont {Silva},
  \citenamefont {Macedo}, \citenamefont {Berti},\ and\ \citenamefont
  {Crispino}}]{Silva:2014fca}%
  \BibitemOpen
  \bibfield  {author} {\bibinfo {author} {\bibfnamefont {H.~O.}\ \bibnamefont
  {Silva}}, \bibinfo {author} {\bibfnamefont {C.~F.~B.}\ \bibnamefont
  {Macedo}}, \bibinfo {author} {\bibfnamefont {E.}~\bibnamefont {Berti}}, \
  and\ \bibinfo {author} {\bibfnamefont {L.~C.~B.}\ \bibnamefont {Crispino}},\
  }\href {\doibase 10.1088/0264-9381/32/14/145008} {\bibfield  {journal}
  {\bibinfo  {journal} {Class.Quant.Grav.}\ }\textbf {\bibinfo {volume} {32}},\
  \bibinfo {pages} {145008} (\bibinfo {year} {2015})},\ \Eprint
  {http://arxiv.org/abs/1411.6286} {arXiv:1411.6286 [gr-qc]} \BibitemShut
  {NoStop}%
\bibitem [{\citenamefont {Doneva}\ and\ \citenamefont
  {Yazadjiev}(2012)}]{Doneva:2012rd}%
  \BibitemOpen
  \bibfield  {author} {\bibinfo {author} {\bibfnamefont {D.~D.}\ \bibnamefont
  {Doneva}}\ and\ \bibinfo {author} {\bibfnamefont {S.~S.}\ \bibnamefont
  {Yazadjiev}},\ }\href {\doibase 10.1103/PhysRevD.85.124023} {\bibfield
  {journal} {\bibinfo  {journal} {Phys.Rev.}\ }\textbf {\bibinfo {volume}
  {D85}},\ \bibinfo {pages} {124023} (\bibinfo {year} {2012})},\ \Eprint
  {http://arxiv.org/abs/1203.3963} {arXiv:1203.3963 [gr-qc]} \BibitemShut
  {NoStop}%
\bibitem [{\citenamefont {{Bowers}}\ and\ \citenamefont
  {{Liang}}(1974)}]{1974ApJ...188..657B}%
  \BibitemOpen
  \bibfield  {author} {\bibinfo {author} {\bibfnamefont {R.~L.}\ \bibnamefont
  {{Bowers}}}\ and\ \bibinfo {author} {\bibfnamefont {E.~P.~T.}\ \bibnamefont
  {{Liang}}},\ }\href {\doibase 10.1086/152760} {\bibfield  {journal} {\bibinfo
   {journal} {\apj}\ }\textbf {\bibinfo {volume} {188}},\ \bibinfo {pages}
  {657} (\bibinfo {year} {1974})}\BibitemShut {NoStop}%
\bibitem [{\citenamefont {{Kippenhahn}}\ and\ \citenamefont
  {{Weigert}}(1990)}]{1990sse..book.....K}%
  \BibitemOpen
  \bibfield  {author} {\bibinfo {author} {\bibfnamefont {R.}~\bibnamefont
  {{Kippenhahn}}}\ and\ \bibinfo {author} {\bibfnamefont {A.}~\bibnamefont
  {{Weigert}}},\ }\href@noop {} {\emph {\bibinfo {title} {Stellar Structure and
  Evolution, XVI, 468 pp.~192 figs..~ Springer-Verlag Berlin Heidelberg New
  York.~Also Astronomy and Astrophysics Library}}}\ (\bibinfo {year}
  {1990})\BibitemShut {NoStop}%
\bibitem [{\citenamefont {Yazadjiev}(2012)}]{Yazadjiev:2011ks}%
  \BibitemOpen
  \bibfield  {author} {\bibinfo {author} {\bibfnamefont {S.}~\bibnamefont
  {Yazadjiev}},\ }\href {\doibase 10.1103/PhysRevD.85.044030} {\bibfield
  {journal} {\bibinfo  {journal} {Phys.Rev.}\ }\textbf {\bibinfo {volume}
  {D85}},\ \bibinfo {pages} {044030} (\bibinfo {year} {2012})},\ \Eprint
  {http://arxiv.org/abs/1111.3536} {arXiv:1111.3536 [gr-qc]} \BibitemShut
  {NoStop}%
\bibitem [{\citenamefont {Sawyer}(1972)}]{Sawyer:1972cq}%
  \BibitemOpen
  \bibfield  {author} {\bibinfo {author} {\bibfnamefont {R.}~\bibnamefont
  {Sawyer}},\ }\href {\doibase 10.1103/PhysRevLett.29.382} {\bibfield
  {journal} {\bibinfo  {journal} {Phys.Rev.Lett.}\ }\textbf {\bibinfo {volume}
  {29}},\ \bibinfo {pages} {382} (\bibinfo {year} {1972})}\BibitemShut
  {NoStop}%
\bibitem [{\citenamefont {Carter}\ and\ \citenamefont
  {Langlois}(1998)}]{Carter:1998rn}%
  \BibitemOpen
  \bibfield  {author} {\bibinfo {author} {\bibfnamefont {B.}~\bibnamefont
  {Carter}}\ and\ \bibinfo {author} {\bibfnamefont {D.}~\bibnamefont
  {Langlois}},\ }\href {\doibase 10.1016/S0550-3213(98)00430-1} {\bibfield
  {journal} {\bibinfo  {journal} {Nucl.Phys.}\ }\textbf {\bibinfo {volume}
  {B531}},\ \bibinfo {pages} {478} (\bibinfo {year} {1998})},\ \Eprint
  {http://arxiv.org/abs/gr-qc/9806024} {arXiv:gr-qc/9806024 [gr-qc]}
  \BibitemShut {NoStop}%
\bibitem [{\citenamefont {{Letelier}}(1980)}]{1980PhRvD..22..807L}%
  \BibitemOpen
  \bibfield  {author} {\bibinfo {author} {\bibfnamefont {P.~S.}\ \bibnamefont
  {{Letelier}}},\ }\href {\doibase 10.1103/PhysRevD.22.807} {\bibfield
  {journal} {\bibinfo  {journal} {\prd}\ }\textbf {\bibinfo {volume} {22}},\
  \bibinfo {pages} {807} (\bibinfo {year} {1980})}\BibitemShut {NoStop}%
\bibitem [{\citenamefont {Thorne}(1980)}]{thorne-MM}%
  \BibitemOpen
  \bibfield  {author} {\bibinfo {author} {\bibfnamefont {K.~S.}\ \bibnamefont
  {Thorne}},\ }\href {\doibase 10.1103/RevModPhys.52.299} {\bibfield  {journal}
  {\bibinfo  {journal} {Rev. Mod. Phys.}\ }\textbf {\bibinfo {volume} {52}},\
  \bibinfo {pages} {299} (\bibinfo {year} {1980})}\BibitemShut {NoStop}%
\bibitem [{\citenamefont {G\"ursel}(1983)}]{Gursel}%
  \BibitemOpen
  \bibfield  {author} {\bibinfo {author} {\bibfnamefont {Y.}~\bibnamefont
  {G\"ursel}},\ }\href@noop {} {\bibfield  {journal} {\bibinfo  {journal} {Gen.
  Rel. Grav.}\ }\textbf {\bibinfo {volume} {15}},\ \bibinfo {pages} {737}
  (\bibinfo {year} {1983})}\BibitemShut {NoStop}%
\bibitem [{\citenamefont {Pappas}\ and\ \citenamefont
  {Apostolatos}(2012)}]{pappas-apostolatos}%
  \BibitemOpen
  \bibfield  {author} {\bibinfo {author} {\bibfnamefont {G.}~\bibnamefont
  {Pappas}}\ and\ \bibinfo {author} {\bibfnamefont {T.~A.}\ \bibnamefont
  {Apostolatos}},\ }\href {\doibase 10.1103/PhysRevLett.108.231104} {\bibfield
  {journal} {\bibinfo  {journal} {Phys.Rev.Lett.}\ }\textbf {\bibinfo {volume}
  {108}},\ \bibinfo {pages} {231104} (\bibinfo {year} {2012})},\ \Eprint
  {http://arxiv.org/abs/1201.6067} {arXiv:1201.6067 [gr-qc]} \BibitemShut
  {NoStop}%
\bibitem [{\citenamefont {Yagi}\ and\ \citenamefont
  {Yunes}(2015{\natexlab{a}})}]{Yagi:2015hda}%
  \BibitemOpen
  \bibfield  {author} {\bibinfo {author} {\bibfnamefont {K.}~\bibnamefont
  {Yagi}}\ and\ \bibinfo {author} {\bibfnamefont {N.}~\bibnamefont {Yunes}},\
  }\href {\doibase 10.1103/PhysRevD.91.123008} {\bibfield  {journal} {\bibinfo
  {journal} {Phys. Rev.}\ }\textbf {\bibinfo {volume} {D91}},\ \bibinfo {pages}
  {123008} (\bibinfo {year} {2015}{\natexlab{a}})},\ \Eprint
  {http://arxiv.org/abs/1503.02726} {arXiv:1503.02726 [gr-qc]} \BibitemShut
  {NoStop}%
\bibitem [{\citenamefont {Yagi}\ and\ \citenamefont
  {Yunes}(2015{\natexlab{b}})}]{in-prep}%
  \BibitemOpen
  \bibfield  {author} {\bibinfo {author} {\bibfnamefont {K.}~\bibnamefont
  {Yagi}}\ and\ \bibinfo {author} {\bibfnamefont {N.}~\bibnamefont {Yunes}},\
  }\href@noop {} {\bibfield  {journal} {\bibinfo  {journal} {in preparation}\ }
  (\bibinfo {year} {2015}{\natexlab{b}})}\BibitemShut {NoStop}%
\bibitem [{\citenamefont {de~Boer}\ \emph {et~al.}(2010)\citenamefont
  {de~Boer}, \citenamefont {Papadodimas},\ and\ \citenamefont
  {Verlinde}}]{deBoer:2009wk}%
  \BibitemOpen
  \bibfield  {author} {\bibinfo {author} {\bibfnamefont {J.}~\bibnamefont
  {de~Boer}}, \bibinfo {author} {\bibfnamefont {K.}~\bibnamefont
  {Papadodimas}}, \ and\ \bibinfo {author} {\bibfnamefont {E.}~\bibnamefont
  {Verlinde}},\ }\href {\doibase 10.1007/JHEP10(2010)020} {\bibfield  {journal}
  {\bibinfo  {journal} {JHEP}\ }\textbf {\bibinfo {volume} {1010}},\ \bibinfo
  {pages} {020} (\bibinfo {year} {2010})},\ \Eprint
  {http://arxiv.org/abs/0907.2695} {arXiv:0907.2695 [hep-th]} \BibitemShut
  {NoStop}%
\bibitem [{\citenamefont {Arsiwalla}\ \emph {et~al.}(2011)\citenamefont
  {Arsiwalla}, \citenamefont {de~Boer}, \citenamefont {Papadodimas},\ and\
  \citenamefont {Verlinde}}]{Arsiwalla:2010bt}%
  \BibitemOpen
  \bibfield  {author} {\bibinfo {author} {\bibfnamefont {X.}~\bibnamefont
  {Arsiwalla}}, \bibinfo {author} {\bibfnamefont {J.}~\bibnamefont {de~Boer}},
  \bibinfo {author} {\bibfnamefont {K.}~\bibnamefont {Papadodimas}}, \ and\
  \bibinfo {author} {\bibfnamefont {E.}~\bibnamefont {Verlinde}},\ }\href
  {\doibase 10.1007/JHEP01(2011)144} {\bibfield  {journal} {\bibinfo  {journal}
  {JHEP}\ }\textbf {\bibinfo {volume} {1101}},\ \bibinfo {pages} {144}
  (\bibinfo {year} {2011})},\ \Eprint {http://arxiv.org/abs/1010.5784}
  {arXiv:1010.5784 [hep-th]} \BibitemShut {NoStop}%
\end{thebibliography}%
\end{document}